\documentclass[two column, pra, showpacs]{revtex4}

\usepackage{graphicx}
\usepackage{dcolumn}
\usepackage{bm}
\usepackage[latin1]{inputenc}

\begin{document}

\preprint{}

\title{Route to turbulence in a trapped Bose-Einstein condensate}

\author{J.~A.~Seman$^1$}\email{jseman@gmail.com}
\author{E.~A.~L. Henn$^1$\footnote{Present address: 5. Physikalisches Institut, Universit\"at Stuttgart, Pfaffenwaldring 57, 70550 Stuttgart, Germany}}
\author{R.~F.~Shiozaki$^1$}
\author{G.~Roati$^2$}
\author{F.~J.~Poveda-Cuevas$^1$}
\author{K.~M.~F.~Magalh\~{a}es$^1$}
\author{V.~I.~Yukalov$^3$}
\author{M. Tsubota$^4$}
\author{M. Kobayashi$^5$}
\author{K. Kasamatsu$^6$}
\author{V.~S.~Bagnato$^1$}

\affiliation{$^1$Instituto de F\'isica de S\~ao Carlos, Universidade de S\~ao Paulo, C.P. 369, 13560-970 S\~ao Carlos, SP, Brazil}
\affiliation{$^2$LENS and Dipartimento di Fisica, Universita di Firenze, and INO-CNR, Via Nello Carrara 1, I-50019 Sesto Fiorentino, Italy}
\affiliation{$^3$Bogolubov Laboratory of Theoretical Physics, Joint Institute for Nuclear Research, Dubna, 141980, Russia}
\affiliation{$^4$Department of Physics, Osaka City University, Sumiyoshi-Ku, Osaka 558-8585, Japan}
\affiliation{$^5$Department of Pure and Applied Science, The University of Tokyo, Komaba 3-8-1, Meguro-ku, Tokyo 153-8902}
\affiliation{$^6$Department of Physics, Kinki University, Higashi-Osaka, Osaka 577-8502, Japan}

\begin{abstract}

We have studied a Bose-Einstein condensate of $^{87}Rb$ atoms under an oscillatory excitation. For a fixed frequency of excitation, we have explored how the values of amplitude and time of excitation must be combined in order to produce quantum turbulence in the condensate. Depending on the combination of these parameters different behaviors are observed in the sample. For the lowest values of time and amplitude of excitation, we observe a bending of the main axis of the cloud. Increasing the amplitude of excitation we observe an increasing number of vortices. The vortex state can evolve into the turbulent regime if the parameters of excitation are driven up to a certain set of combinations. If the value of the parameters of these combinations is exceeded, all vorticity disappears and the condensate enters into a different regime which we have identified as the granular phase. Our results are summarized in a diagram of amplitude versus time of excitation in which the different structures can be identified. We also present numerical simulations of the Gross-Pitaevskii equation which support our observations.

\end{abstract}


\maketitle

\section{Introduction}

Turbulence is a phenomenon well known for classical fluids. Superfluids also exhibit this phenomenon \cite{Feynman,Hall,Vinen1,Vinen2,Vinen3,Vinen4} and in this case it is known as Quantum Turbulence (QT). It consists of a collection of quantized vortices tangled in space \cite{Donnelly}. In the case of superfluid Helium, QT has been achieved in many ways, for instance in thermal counterflow driven by heat injection \cite{Vinen1,Vinen2,Vinen3,Vinen4} or by vibrating objects within the liquid \cite{QTexp1,QTexp2,QTexp3}.

It is believed that experiments with superfluids may overcome many of the complications encountered in classical fluids \cite{Vinen1,Vinen2,Vinen3,Vinen4}. In this sense, the existence of quantized vortices in quantum fluids may simplify the interpretation and provide grounds for a theoretical description of the problem. However, due to the high density and strong interactions present in superfluid Helium, the size of the vortex cores is very small. As a consequence, the observation of QT and its dynamics is indirect and the interpretation of the experiments is difficult \cite{QTexp1,QTexp2,QTexp3}

Trapped Bose-Einstein condensates represent an ideal system for studying QT. For being dilute and weakly interacting systems, the vortices can be directly observed by optical absorption. Additionally, because of the small scale and the high controllability that these systems offer, QT can be generated and studied under very controlled conditions. Therefore, the emergence of QT in atomic superfluids may create new and exciting alternatives for exploring this phenomenon \cite{PhysToday}.

Recently, QT was detected in a BEC of $^{87}Rb$ atoms in a magnetic trap \cite{PRL}. The turbulent state was achieved by introducing an external excitation through an oscillatory magnetic field. In a subsequent publication it has been discussed how the emergence of QT is related to the finite size of the atomic condensates \cite{LPL}. The finiteness of the system is a particular property of ultracold trapped gases not present in superfluid helium. Therefore, even though QT is the same phenomenon in both BECs and superfluid Helium, the reason for its appearance in these systems can be very different.

In the present letter we offer new studies in which we show how the variation of the parameters of the excitation must be performed in order to generate very different regimes in the sample. Specifically, we vary the time and amplitude of the excitation and observe the effect produced in the sample. Depending on the combination of the time and amplitude, the excitation is able to generate four different regimes, namely: (\textit{i}) the bending of the main axis of the cloud, (\textit{ii}) the nucleation of regular vortices, (\textit{iii}) the quantum turbulent regime, and (\textit{iv}) the granular phase of the system, whose observation we report for the first time. Our results are summarized in a diagram of amplitude versus time of excitation in which the four different regimes can be identified. This diagram represents a very novel result because, due to the finiteness of the atomic superfluids, it is exclusive of BECs and, certainly, it is not present in superfluid Helium. The diagram is also important because it shows the route in which the parameters of excitation must be combined to reach the different observed states.

Of particular interest is the way in which the array of vortices evolves to the turbulent regime. For explaining this transition we have performed a numerical simulation of the Gross-Pitaevskii equation, obtaining qualitative agreement with the observations. Finally, we offer a plausible explanation for the granular phase and discuss future directions for its understanding.

\section{Experimental system}

The experimental setup to produce the BEC and to excite it up to the turbulent regime is described in detail in our previous publications \cite{PRL,JLTP}. In brief, we produce a cigar-shaped BEC of $^{87}Rb$ with $(2-3)\times10^5$ atoms in a harmonic magnetic trap with frequencies $\omega_x  = 2 \pi\times 23$~Hz and $\omega_r = 2 \pi \times 210$~Hz. Once the condensate is obtained an oscillating magnetic field is superimposed to the trapping potential. This oscillation is followed by an equilibrating time of 20~ms before releasing the atoms for free expansion.

The oscillatory field has a fixed frequency of $2\pi \times 200$~Hz and is produced by a pair of anti-Helmholtz coils. This excitation field generates distortions on the potential, promoting a combination of deformation, displacement, and rotation in the cloud. As shown in references \cite{PRL,LPL,JLTP,PRA}, the excitation is able to nucleate vortices in the sample and, for the proper conditions, to take it to the turbulent regime.

The possibility of generating strongly nonequilibrium BECs with vortices and other coherent topological modes by modulating the trapping potential was discussed earlier 
\cite{YYB97,YYB02,YB09}. The prevailing creation of vortices by such a modulation owes to the fact that vortices with unit circulation are the most stable among all other topological modes. More precisely, the trap modulation, imposing no total circulation on the atomic cloud, creates vortex-antivortex pairs randomly distributed in space. This is different from the creation of vortices by an effective condensate rotation \cite{PS08,KP09}.   

\section{Oscillatory excitations and produced states}

\begin{figure}
\centering
 \includegraphics[scale=0.29]{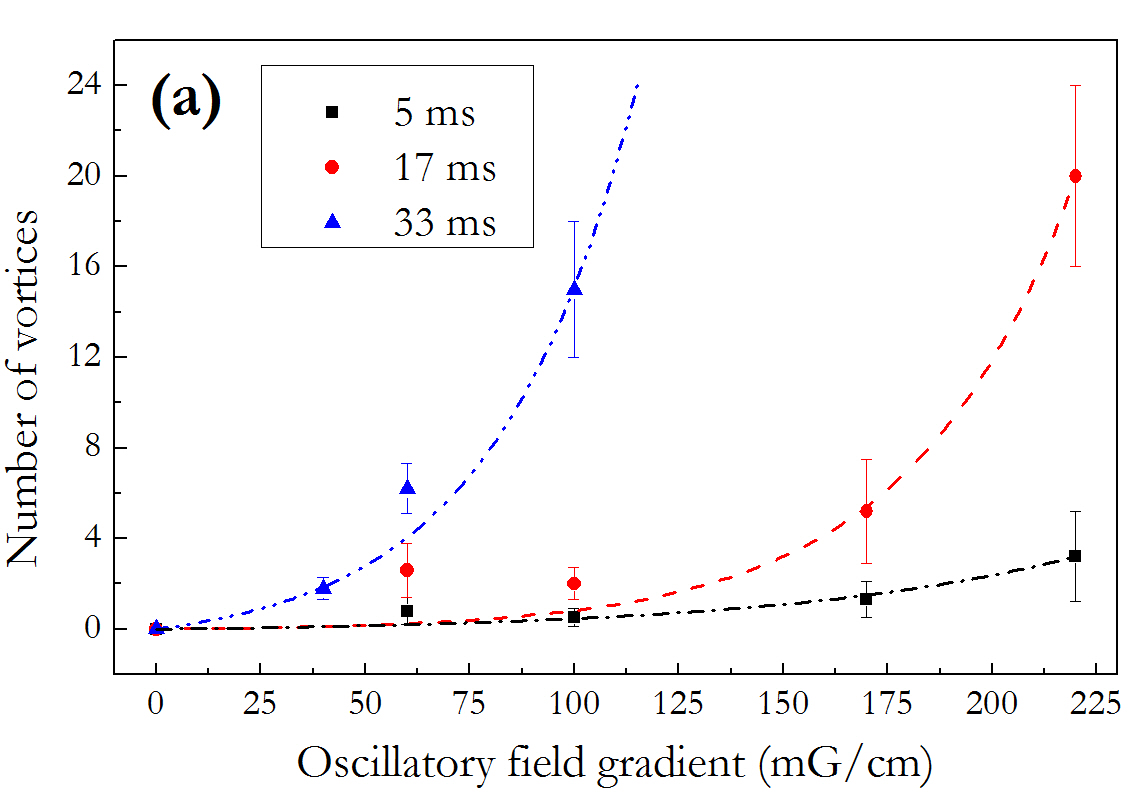}
 \includegraphics[scale=0.30]{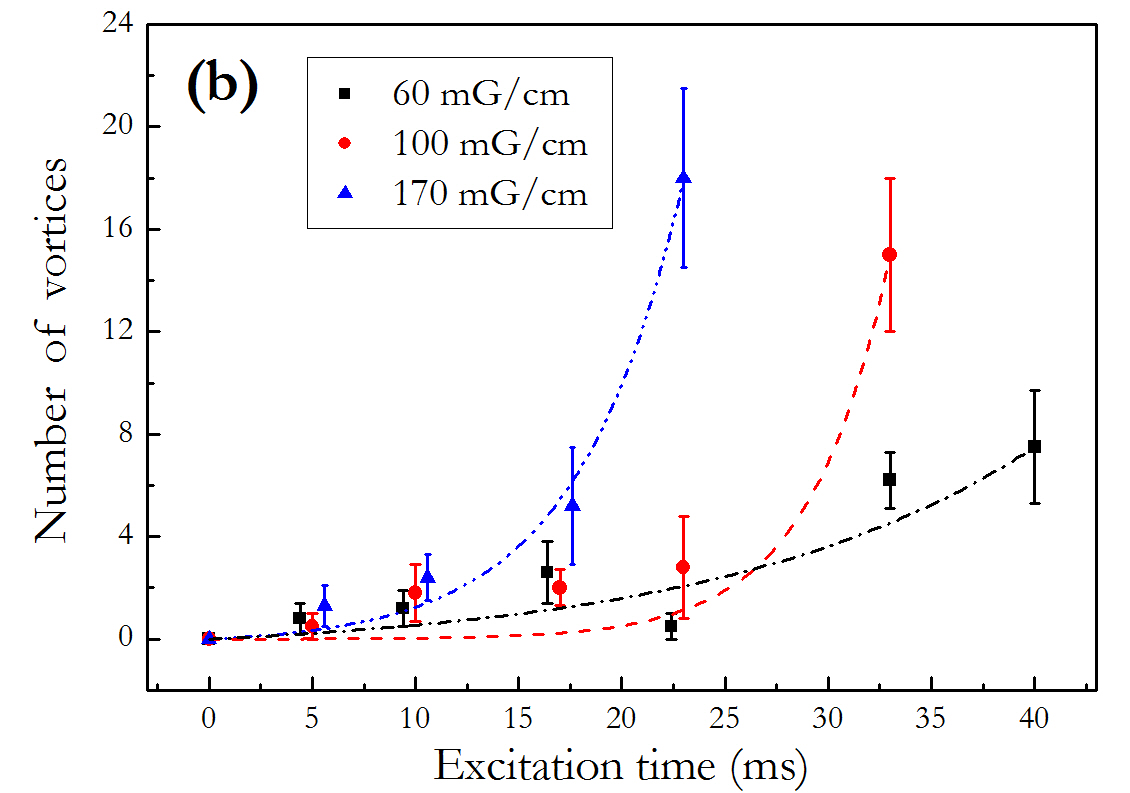}
 \caption{Average number of vortices observed in the excited cloud as a function of (a) the excitation amplitude for three different excitation times and (b) as a function of the excitation time for three different amplitudes of excitation. Lines are guides for eyes. The error bars show the standard deviation of the mean value of the number of vortices.}
\label{fig:time}
\end{figure}

We have considered a range of combinations of excitation times and amplitudes. For small amplitudes, independently of the excitation time, the only effect on the cloud is the bending of its main axis. That result was reported previously and it is an inherent consequence of the superfluid nature of the atomic cloud \cite{scissors1,scissors2}. Increasing the amplitude leads to a monotonic increase of the average number of formed vortices, although there is a big variation in the number of vortices when the same conditions are employed. The results for three different excitation times are presented in Fig.~\ref{fig:time}(a).

The distribution of vortices in our sample does not correspond to any regular pattern similar to those reported in the studies related with the formation of vortex lattices \cite{Lattice1,Lattice2}. We believe that this is a consequence of our vortex formation mechanism, in which the nucleation of vortices and anti-vortices is equally likely. In fact, considering configurations of three vortices, we have collected evidence that it is vortices and anti-vortices that are formed in the cloud \cite{3vortex}. Even in the absence of regularity, the states where the number of vortices in the sample can be well identified will be called {\it regular}.

As the parameters of the excitation increase, when a number of vortices of the order of 20 is reached, a change in the behavior of the distribution of vortices in the cloud is observed. This corresponds to the formation of a random vortex tangle, one of the main features of QT. These distributions exhibit a large shot-to-shot variability. Additionally, the turbulent cloud demonstrates isotropic behavior in its free expansion, as previously described \cite{PRL}.

The graph of Fig.~\ref{fig:time}(a) shows a clear connection between the time and the amplitude of the excitation. Excitations for an extended period of time reach the turbulent condition for much lower amplitudes. Alternatively, the amplitude can be fixed and the time of excitation varied. Fig.~\ref{fig:time}(b) shows the average number of vortices as a function of the excitation time for three fixed amplitudes. While no big difference is observed for the shorter excitation times, at longer times, the larger amplitude forces the system to the divergence in the number of vortices and, consequently, to the turbulent regime.

After reaching QT, increasing even further the time and/or the amplitude of excitation renders the sample to a complete granulation, where the pieces of condensate are spread out over the cloud. The grains persist if the  pumping of energy is kept flowing into the sample. The granular state of BEC is a kind of a heterophase mixture where BEC droplets are surrounded by normal gas \cite{Y10}. Experimental confirmation concerning the superfluidity of each grain still remains to be addressed.

\begin{figure}
\centering
 \includegraphics[scale=0.70]{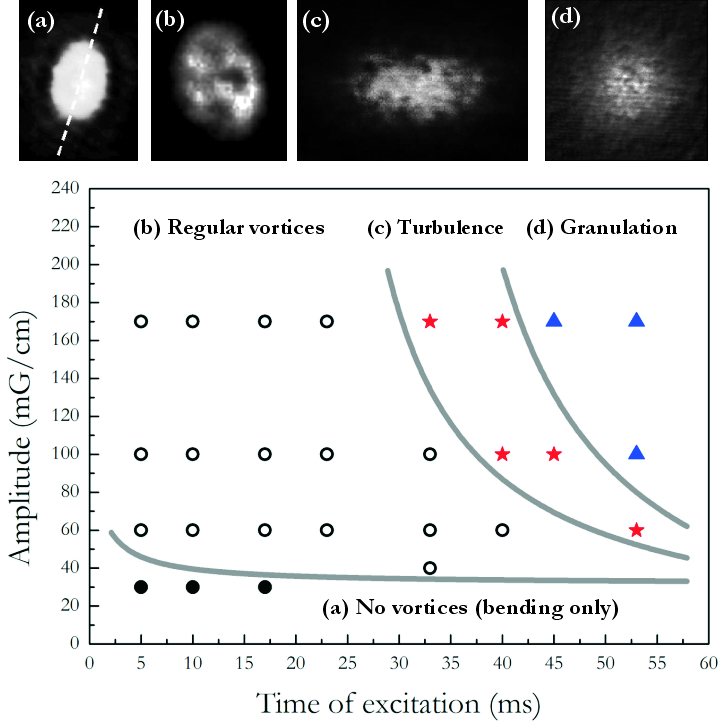}
 \caption{Diagram showing the domains of the parameters associated with the observed regimes of the atomic superfluid. The points are the experimental observations. Figures on the top correspond to typical observations. For the region of regular vortices, the number of vortices varies with the parameters as presented in Figs.~\ref{fig:time}(a) and \ref{fig:time}(b). Gray lines are guides for eyes, separating the domains of different observations.}
\label{fig:diagram}
\end{figure}

The diagram of Fig.~\ref{fig:diagram} shows the correspondence between the observed distributions and the combinations of time and amplitude of excitation. It exhibits four domains, each corresponding to one of the observed regimes. The border lines between these domains are just guides to eyes. This diagram is useful for understanding the route to the turbulent regime. It shows that the important quantity related to this route is the product of the amplitude and excitation time. This is equivalent to considering the total energy pumped into the cloud and could be used for characterizing the threshold behavior. In fact, the number of formed vortices is related to the total energy pumped into the system. The threshold for reaching turbulence also must depend on the size of the cloud, as discussed in Ref.~\cite{LPL}. Since QT occurs when the number of vortices is around 20, one should expect that the cloud must be sufficiently large to contain many vortices.

\section{Numerical calculations for turbulent regime}

For better understanding the experimental results, we have performed numerical simulations based on the Gross-Pitaevskii equation, with the addition of a phenomenological term, $\gamma$ to account for dissipation. In dimensionless units, the equation for our simulations is given by \cite{Choi,Tsubota}
\begin{eqnarray}\label{eq:GPE}
\left(i-\gamma\right)\frac{\partial\Psi}{\partial t} = \left(-\frac{\nabla^2}{2} + V -\mu + u\left|\Psi\right|^2\right)\Psi
\end{eqnarray}

The net potential acting on the atoms is the sum of the harmonic magnetic trap and the oscillatory field, and can be approximately expressed by
\begin{eqnarray}\nonumber\label{eq:V}
V & = & \frac{1}{2}\left\{\lambda^2\left[x\cos\theta_1 + y\sin\theta_1\right.\right. \\
  & & \left. - z\sin\theta_2 - \delta_1\left(1 - \cos\Omega_0 t\right)\right]^2 \\ \nonumber
  & & + \left[y\cos\theta_1 - x\sin\theta_1 - \delta_2\left(1-\cos\Omega_0 t\right)\right]^2 \\ \nonumber
  & & \left.+ \left[z\cos\theta_2 + x\sin\theta_2 - \delta_3\left(1 - \cos\Omega_0 t\right)\right]^2 \right\}, 
\end{eqnarray}
where $\lambda = \omega_x/\omega_r$ and $\theta_i = A_i\left(1 - \cos\Omega_0 t\right)$ are time dependent angles. For our experimental conditions, $\Omega_0 = 2\pi\times 200$~Hz, $A_1 \simeq \pi/60$ and $A_2 \simeq \pi/120$. The amplitudes for the translational oscillation of the potential minimum are $\left( \delta_1, \delta_2, \delta_3 \right) = \alpha\left( 2, 5, 3 \right)\mu \mbox{m}/a_r$, where $a_r = \sqrt{\hbar/m\omega_r}$ and $\alpha$ is a variable parameter that represents the amplitude of the center-of-mass oscillation. Hence, the parameter $\alpha$ is proportional to the amplitude of the excitation. 
We employ the representation $\Psi = \psi(y, z)\phi(x)$ in Eq.~(\ref{eq:GPE}) and consider $2D$ simulations in the $y-z$ space. In this case, the interaction term becomes $u_{2D} = 4\pi aN/R_x$, with $R_x$ being the characteristic size of the condensate along $x-$axis. For our experiment $u_{2D} = 740$. Since the thermal atoms, which cause the dissipation, move together with the potential, we also have to consider the reference frame co-moving with the potential. In this frame Eq.~(\ref{eq:GPE}) becomes
\begin{eqnarray}\nonumber
\left(i-\gamma\right)\frac{\partial\psi}{\partial t} = \left(-\frac{\nabla^2}{2} + V_0 -\mu + u_{2D}\left|\psi\right|^2 \right.\\
\left.-\bf{\Omega}(t)\cdot\hat{\bf{L}} - \bf{v}(t)\cdot\hat{\bf{p}}\right)\psi, \label{eq:refGPE}
\end{eqnarray}
with the linear momentum $\hat{\bf p} = - i \nabla$ and the angular momentum $\hat {\bf L} = \hat {\bf r} \times \hat{\bf p} $. We consider $V_{0} = (y^2 + z^2) / 2$,  ${\bf v}(t) =  (0, v_{y}, 0) \sin \Omega_{0} t $ and ${\bf \Omega} (t) = (\Omega_x, 0, 0) \sin \Omega_{0} t $. Using the half of the oscillation period $T  = \pi/ \Omega_{0}$, we obtain $ v_{y} \simeq 2 \delta_{2} / T = 2 \Omega_0 \delta_{2} / \pi $. The rotation frequency $\Omega_{x}$ is also estimated as $\Omega_{x} \simeq 2 A_{2} / T = \Omega_0 / 60$, providing a very small contribution.

\begin{figure}
\centering
 \includegraphics[scale=0.12]{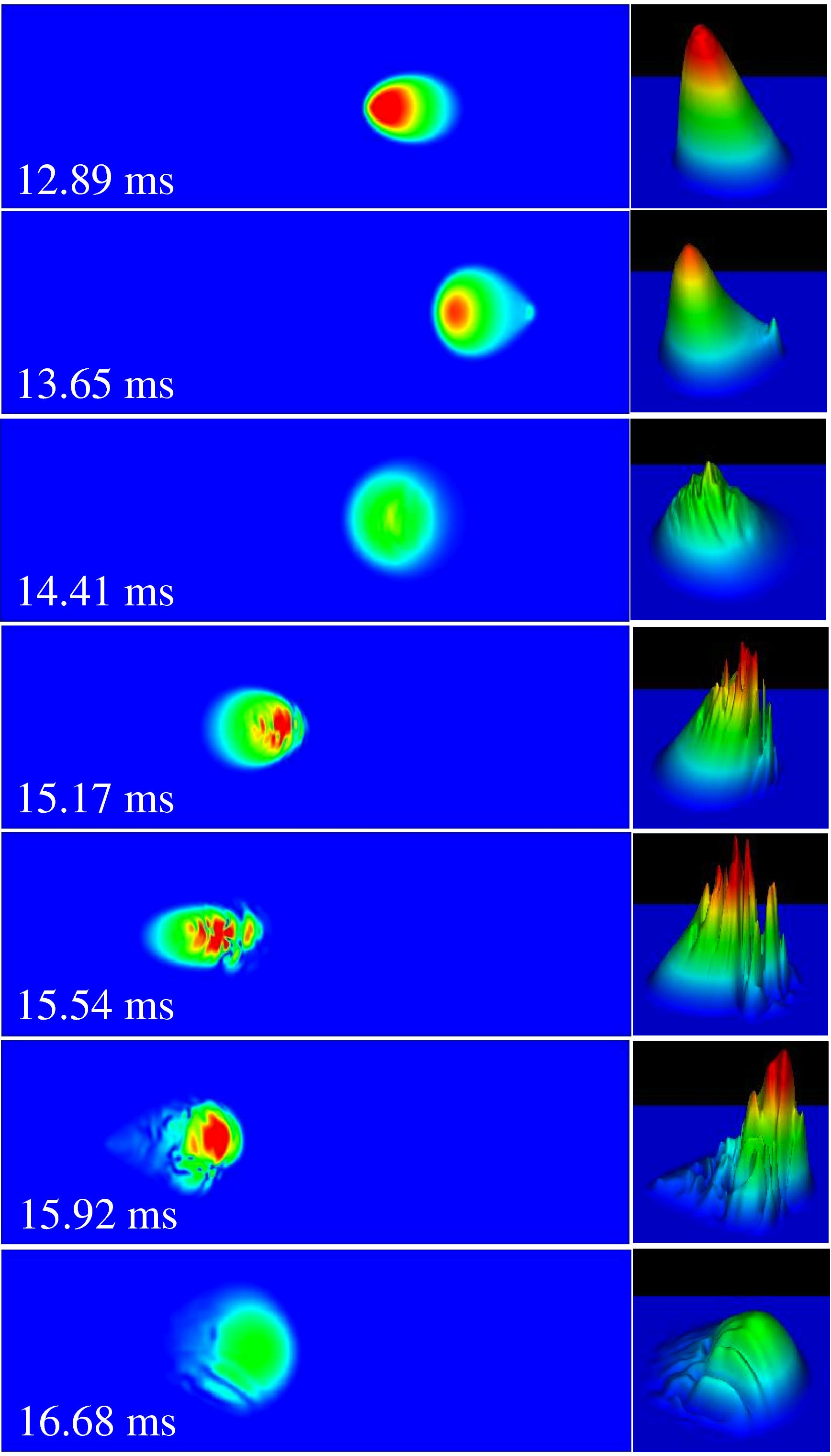}
 \caption{Snapshots of the BEC after different times of excitation. The left and the right columns show the 2D and 3D plots of the density profile, respectively. The colors range from red (high density) to blue (low density).}
\label{fig:snap}
\end{figure}

Fig.~\ref{fig:snap} shows snapshots of the density profile for different excitation times ranging from 13 to 17~ms. Additionally, we have calculated the mean angular momentum per atom, $\left\langle L_x\right\rangle = \int d\bf{r}\psi^*\hat{L}_x\psi$, as a function of the excitation time. Using $\alpha = 1.6$ and $\gamma = 0.02$, our simulations show that $\left\langle L_x\right\rangle$ blows up after $15$~ms of excitation. At this point, the condensate forms wavy patterns which develop to dark solitary waves which subsequently decay into several vortex pairs via the snake instability \cite{Feder}. A more complex dynamics takes place after the first events of vortex formation, consisting in the generation of an undetermined number of vortices which characterizes the emergence of the turbulent regime. 

As $\alpha$ increases the nucleation of vortices occurs at earlier times, with a faster evolution to QT. This agrees well with the observations presented in the diagram of Fig.~\ref{fig:diagram}. It is important to mention that the time scale of the vortex events of the simulations, which is of the order of 10~ms, is consistent with the times observed in the experiment.

These results demonstrate that the combination of rotation and translation is essential to produce vortices with lifetimes long enough to be observed after time-of-flight imaging. The instability on vortex generation may be associated with the Landau's argument for the critical velocity \cite{Landau} above which a fluid becomes turbulent. However, we must recall that, in general, the creation of vortices is not necessarily associated with the Landau critical velocity, as discussed in Ref.~\cite{Berloff}.

The dynamics of the BEC strongly depends on the strength of dissipation $\gamma$. If the dissipation is absent, no instability associated with the soliton creation occurs. Values of $\gamma$ between $0.015$ and $0.025$ are optimal for the generation of vortices and quantum turbulence. The simulations cannot reproduce the whole observed diagram, since our system is a 3D gas. Nevertheless, good qualitative agreement with the experiment has been achieved.

\section{Granular state}

When an overdose of energy is pumped into the turbulent cloud, the condensate enters in a new regime where all vorticity disappears and the system exhibits a new phase. In this case, the superfluid system demonstrates a kind of burst into grains. A typical image of this state can be seen in the insert (d) of Fig.~\ref{fig:diagram}. The resulting state is analogous to the Bose glass state, where the grains of condensate are surrounded by uncondensed atoms. The origin of this granulation can be understood as follows. The action of an external alternating field can be shown to be equivalent, on average, to the action of an external spatially random potential \cite{Yukalov}. For an equilibrium system, the granular condensate appears under the increasing amplitude of the external spatially random potential \cite{Y10,Y09}. Similarly, for a nonequilibrium system, subject to the action of an alternating field, the granular state arises under the increasing value of the energy pumped into the system. The relation between the value of the oscillatory field amplitude, $V_0$, and the effective trap frequency $\omega_0$ is crucial for the occurrence of the corresponding states. This is connected with the relation between the characteristic trap length $l_0 = 1/\sqrt{m\omega_0}$, playing the role of the correlation length, and the localization length $l_{loc} \sim (\omega_0/V_0)^2l_0$, being an analog of the Larkin length \cite{Larkin}.

When the energy pumped by the alternating potential is small ($V_0 \ll \omega_0$), there is an extended condensate filling the trap. This condensate can be strongly turbulent, but it still fills the whole trap. If the amplitude or modulation time of the alternating potential is increased, the condensate becomes strongly perturbed, eventually reaching the point where $l_{loc} \sim l_0$. Under this condition, the condensate cannot be sustained as a whole and fragments into pieces. The nonequilibrium granular condensate can exist for the range $\omega_0 \leq V_0 \leq \omega_0\sqrt{l_0/a}$. When the pumped energy reaches the upper boundary, $V_0 \sim \omega_0\sqrt{l_0/a}$, then $l_{loc} \sim a$ and the granular condensate should be completely destroyed. This boundary was not reached in our experiment. Our observations are in agreement with the expected average size and number of the condensate droplets in the granular phase.

The state, requiring very strong pumping and appearing after the granular condensate is destroyed, can correspond to nonequilibrium normal fluid in a chaotic regime \cite{Y10}. This state can be similar to \emph{weak turbulence}. The latter, as discussed in \cite{Berloff2}, is not related to vorticity but is characterized by strong fluctuations of the density distribution. The evolution from the granular to weak turbulent regime may be possible in the present system, which however, requires further experimental investigations. Another interesting direction for experimental studies could be the process of relaxation of the nonequilibrium trapped system after the end of external perturbations. Finite quantum systems can demonstrate a number of peculiarities in the process of their equilibration \cite{PSSV10,Y11}.

\section{Conclusions}

Turbulence presents one of the most challenging phenomena in physics. Even though this field has attracted high attention, its understanding is yet far from being perfect and a complete description of the problem is still missing. When turbulence occurs in superfluids it is known as Quantum Turbulence (QT) and presents many features which have no equivalent in its classical counterpart. In particular, the fact that the vortices are quantized makes the description of the turbulent flow simpler than in classical fluids. For this reason, it is believed that superfluids are excellent model systems for understanding such an important topic. To date, most of the work involving QT has been focused on the case of superfluid Helium (He-4 and He-3).
Nevertheless, the possibility of studying this phenomenon in a Bose-Einstein condensate (BEC) opens up new and exciting research opportunities mainly for three reasons which are not found in the case of Helium:

(i) Condensates offer a very high degree of controllability. In a BEC, it is possible to control the number of particles, interaction strength, and the confining potential where the atoms are trapped, making it an ideal system for generating QT in a highly controllable manner.

(ii) Trapped atoms form weakly interacting systems, where the vortex cores are essentially larger than in superfluid Helium, making their observation much easier.

(iii) The fact that the BEC is a finite-size system presents a new scenario for discovering new effects associated with QT.

In the present work we performed new studies of QT. In particular, we described the routes by which QT can be generated by means of an oscillatory excitation. We studied the evolution of a non-turbulent system to a turbulent one by varying the parameters of an oscillatory excitation applied to the sample. These observations, although peculiar for our system, are very important allowing for the general understanding of the conditions under which this phenomenon can be produced and investigated. Besides the turbulent regime, when the high values of the parameters are employed, we also observe, for the first time, a different regime, which we have identified as granulation of the BEC. In this case, the excitation is so high that the condensate breaks into small pieces surrounded by uncondensed atomic clouds. We summarized our observations in a diagram of excitation parameters, identifying the domains of different nonequilibrium states and clarifying the route by which these regimes could be reached. This diagram, being principally novel, is peculiar of trapped atomic systems, whose size is finite and, therefore, their behavior can be drastically different from that of bulk superfluids, such as superfluid Helium. Together with our experimental observations, we also provide a numerical model supporting our findings and allowing for a qualitative understanding of the evolution to quantum turbulence. We offered a plausible explanation of the arising granulated phase and suggested ways for its further investigation.

The main results of the paper can be briefly summarized as follows. We have constructed a diagram showing the regions of different states realized under the combined variation of the time and amplitude of an oscillatory excitation. The gradual evolution from a bended BEC to the regular vortex state, to the turbulent regime, and, finally, to granulated condensate is presented. Numerical simulations allow us to qualitatively explain the observations and to identify the requirements for realizing this or that regime. The obtained diagram serves as a guide demonstrating the parameters that are necessary for experimentally producing different nontrivial nonequilibrium states of trapped atoms, such as turbulent condensates and granular condensates.

\section*{ACKNOWLEDGMENTS}

We appreciate financial support from the Brazilian agencies Fapesp and CNPq, the Russian Foundation for Basic Research, and Grant-in-Aid for Scientific Research from JSPS.

\end{document}